\documentclass{ws-mpla}

\begin{document}

\markboth{D. Lacroix, G. Hupin}
{Density functional for pairing with particle number conservation}

\catchline{}{}{}{}{}

\title{{ \bf DENSITY FUNCTIONAL FOR PAIRING WITH PARTICLE NUMBER CONSERVATION}
}

\author{\footnotesize DENIS LACROIX\footnote{lacroix@ganil.fr} ~and GUILLAUME HUPIN}
\address{GANIL, CEA and IN2P3, Bo\^ite Postale 55027, \\
14076 Caen Cedex, France}

\maketitle


\begin{abstract}

In this work, a new functional is introduced to treat pairing correlations in finite many-body systems.
Guided by the  projected BCS framework, the energy is written as a functional of occupation 
numbers. It is shown to generalize the BCS approach and to provide an alternative to Variation After Projection 
framework. Illustrations of the new approach are given for the pairing Hamiltonian for various particle numbers 
and coupling strengths. In all case, a very good agreement with the exact solution is found.
\keywords{pairing; mesoscopic systems; particle number projection}
\end{abstract}

\ccode{PACS Nos.: 74.78.Na , 21.60.Fw , 71.15.Mb}

\section{Introduction}	

\begin{figure}[htbq]
\begin{center}
\includegraphics[width = 5.cm]{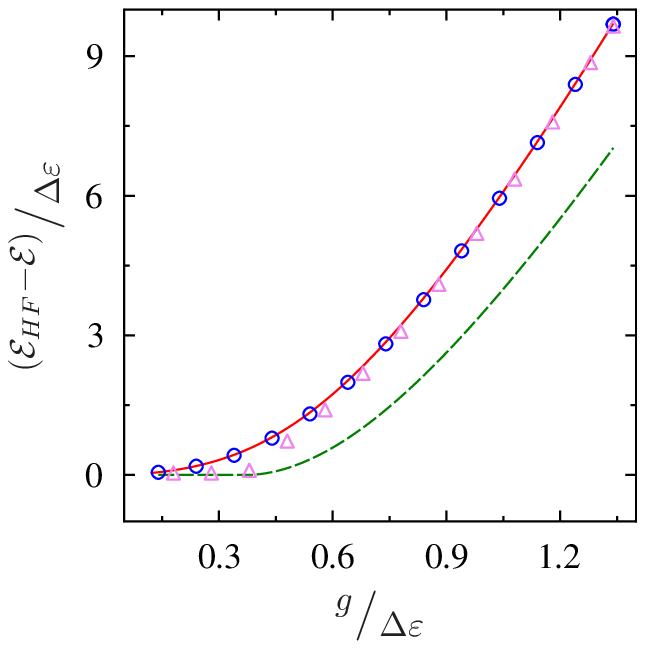} \hspace*{0.5cm}
\includegraphics[width = 5.cm]{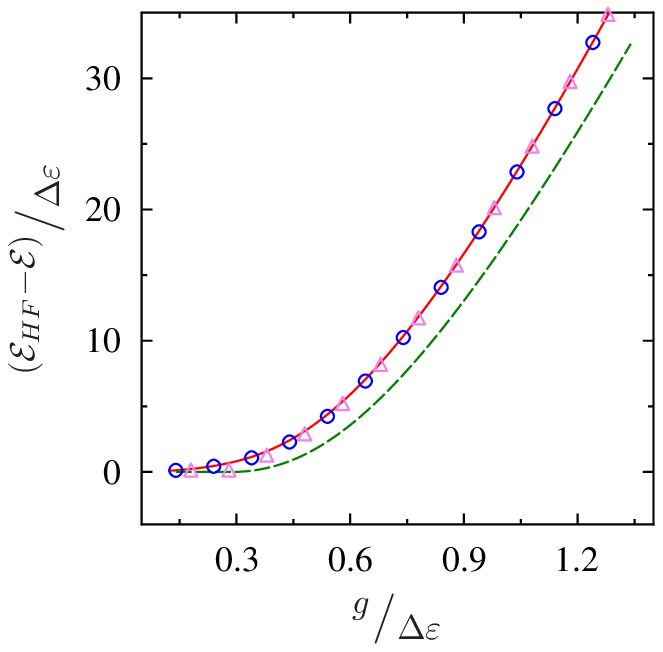}
\end{center}
\caption{ \label{fig1:niigata} Difference between the exact total energy and the Hartree-Fock energy  (solid line) 
obtained in the picket fence Hamiltonian with constant level spacing $\Delta \varepsilon$ for $A=8$ and $A=16$ particles. In both 
case, the BCS (dashed), PAV (open triangle) and VAP (open circles) results are displayed.}
\end{figure}

The Nuclear Energy Density Functional (EDF) theory is anticipated to provide 
a unified framework to describe nuclear structure and reactions. Important 
efforts are now being made to improve the predicting power of EDF.  An important aspect 
of current EDF is the possibility to break some of the symmetries of the nuclear Many-Body
problem in order to grasp specific correlations with rather simple functionals. A typical example is provided 
by pairing correlations that are approximately treated by relaxing the particle number conservation, like 
in the BCS or HFB theory. Such theories are expected to be rather effective in the large particle number
limit but miss important effects as the particle number  and/or  coupling decreases. Projection 
techniques that restore the particle number  are expected to significantly improve the description of pairing  in
that cases\cite{Rin80}. Figure \ref{fig1:niigata} illustrates the energy for the "picket fence"
pairing Hamiltonian \cite{Ric64}, 
\begin{eqnarray}
H = \sum_{i} \varepsilon_i (a^\dagger_i a_i + a^\dagger_{\bar i} a_{\bar i}) - g \sum_{i,j }  a^\dagger_i 
a^\dagger_{\bar i} a_{\bar j} a_{j}, \label{eq:hrich}
\end{eqnarray} 
obtained in the BCS, and projected BCS [PBCS] frameworks  when the variation is
made before (Variation After Projection [VAP]) or after the variation (Projection After Variation [PAV]). 
The projection clearly improves the description of correlation and is almost indistinguishable from 
the exact result when VAP is performed.

Projection on particle number and/or angular momentum is becoming now a standard tool 
of nuclear EDF and extensive calculations along the whole nuclear chart is within reach. Recent 
discussions have however pointed out that projection should be handled with care when combined 
with functional theory\cite{Ang01,Dob07}.  In particular, specific correction to the so-called self-interaction problem 
in "GCM like" calculations are necessary\cite{Lac09,Ben09,Dug09}. 
In addition, even with proper corrections, projections especially
made before the variation\cite{She00} remain very heavy numerically and their use for other problems than nuclear structure,  
like for instance the dynamics or thermodynamics of nuclear systems, seems difficult. 
The aim of the present work is to propose a new functional for pairing
able to directly grasp correlations generally incorporated through projection by specific dependence 
on the the natural occupation numbers $n_i$ and orbitals $\{ \varphi_i \}$, i.e. 
$\rho=\sum_i | \varphi_i \rangle n_i \langle \varphi_i |$ where $\rho$ denotes 
the exact one body density matrix.  

\section{Density Matrix Functional for pairing}

Our starting point is the projected BCS state given by:
\begin{eqnarray}
| N \rangle & \equiv &  \frac{1}{\sqrt{N!}} \left( \sum_i x_i a^\dagger_i a^\dagger_{\bar i}  \right)^N | - \rangle, \label{eq:pbcsstate}
\end{eqnarray}
where $\{a^\dagger_i  , a^\dagger_{\bar i} \}$ denotes pairs of time-reversed states.   It could
easily be shown that this state can be obtained by projection of the BCS state, 
$| {\rm BCS} \rangle = \prod(1+x_i a^\dagger_i a^\dagger_{\bar i})| 0 \rangle | - \rangle$, onto good particle number $A=2N$. Note that here $N$ denotes the number of pairs. The PBCS 
energy for the Hamiltonian (\ref{eq:hrich}) reads
\begin{eqnarray}
\frac{\langle N|H | N \rangle }{\langle N | N \rangle} \equiv {\cal E} (n_i , C_{ij}) =  2 \sum_i  \varepsilon_i n_i - g  \sum_{ij} C_{ij} \label{eq:nicij},
\label{eq:nici}
\end{eqnarray}
where $n_i$ and $C_{ij}$ are the occupation probabilities and correlation matrix elements defined 
through 
\begin{eqnarray}
n_i &=& \frac{\langle N |a^\dagger_i a_i | N \rangle }{\langle N | N \rangle }, ~~ 
C_{ij} = \frac{\langle N |b^\dagger_i b_j | N \rangle }{\langle N | N \rangle }.
\end{eqnarray}
According to the definition (\ref{eq:pbcsstate}), $n_i$ and $C_{ij}$ and therefore $ {\cal E} (n_i , C_{ij})$ can eventually 
be written as an explicit functional of the $\{x_i \}$ parameters, which turns out to be too complicated for a direct use as 
variational parameters.  After tedious manipulation, it has been shown in ref. \cite{Lac10}, that these parameters can 
approximately be written as a functional of the $\{ n_i \}$ through:
\begin{eqnarray}
|x_i|^2 &\simeq& \left(\frac{n_i}{ 1 -n_i} \right)  (a_0 - a_1 n_i)
\end{eqnarray}
with 
\begin{eqnarray}
a_1 &=&    \frac{1}{N} \left( 1+  s_2 + s_2^2  + 
\cdots + s_2^{N-1}   \right) =   \frac{1}{N}  \frac{1 - s_2^N}{1 - s_2} 
\end{eqnarray}  
and 
\begin{eqnarray}
a_0 &=&    1 + \frac{(s_2-s_3)}{N}  
 \left( 1+ 2  s_2 + \cdots +(N-1)   s_2^{N-2}   \right) 
       = 1  + (s_2-s_3)  \frac{\partial a_1}{\partial s_2} , \label{eq:a0a1}
\end{eqnarray}  
and $s_p = 1/N \sum n^p_i$. With these expressions, we can now write the correlation as a functional of single-particle 
occupancies:
\begin{eqnarray}
C_{i  j } &=&  \sqrt{n_i (1-n_i) n_j (1-n_j) } \times \frac{ \sqrt{ (a_0 - a_1 n_i)(a_0 - a_1 n_j)}}
{\left\{ a_0 - a_1 (n_i + n_j - n_i n_j )  \right\} }. \label{eq:cijfunc}
\end{eqnarray}  
The present functional, based on the PBCS trial state, corresponds to a generalization of the BCS ansatz that 
approximately account for particle number conservation. The BCS limit is obtained for $a_0=1$ and $a_1 = 0$ leading to
the well know expression $C_{i  j } = \sqrt{n_i (1-n_i) n_j (1-n_j) }$. It is worth to mention that a completely 
different expression is obtained in the weak coupling limit (Hartree-Fock limit) for which $s_2 = s_3 = 1$ and 
$C_{i  j } \rightarrow  \sqrt{n_i n_j}$.

\begin{figure}[htbq]
\begin{center}
\includegraphics[width = 5.cm]{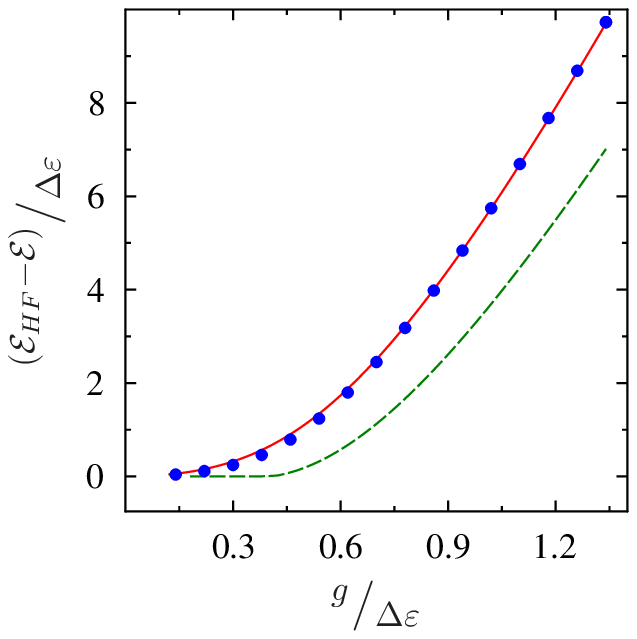} \hspace*{0.5cm}
\includegraphics[width = 5.cm]{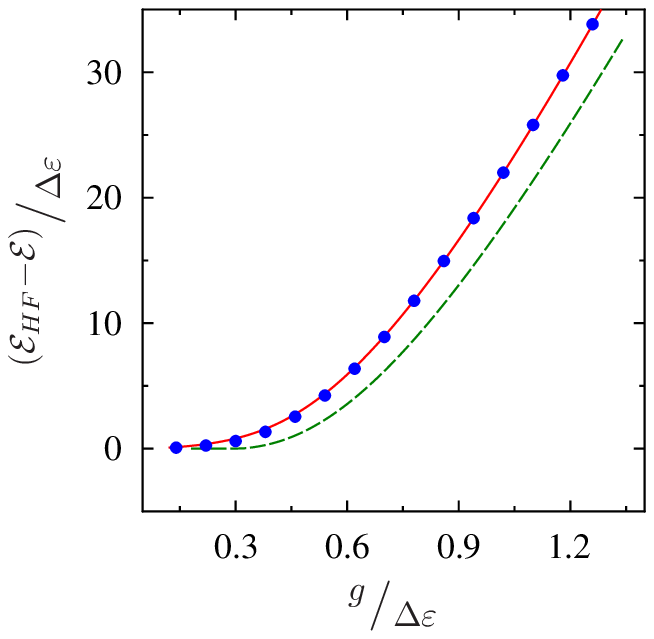}
\end{center}
\caption{ \label{fig2:niigata} Same quantity as in Figure \ref{fig1:niigata} displayed as a function of the coupling strength 
for the exact (solid line), BCS (dashed line) and the new functional (filled circles) case.}
\end{figure}

Illustrations of the new functional accuracy are given in figure \ref{fig2:niigata} where the energy is 
obtained by a direct minimization of (\ref{eq:nicij}) using expression (\ref{eq:cijfunc}).  
A very good agreement is obtained for any particle number and coupling strength.
  
\section{Summary}

Following the Density Matrix Functional Theory spirit \cite{Gil75}, the possibility to use functionals 
of natural orbitals and occupancies has been recently introduced in nuclear physics \cite{Pap07,Lac09b}.
In the present work, a new functional for pairing is introduced that correct the BCS theory for particle number 
conservation effects. The present framework provide an alternative way to perform VAP calculations and is expected 
(i) to greatly simplify such calculation compared to direct use of projection operator,
(ii) to directly  give access to physical quantities like natural occupancies,
(iii) to be particularly suitable for EDF based theories.

\end{document}